\documentclass[%
 reprint,
 showkeys,
 amsmath,amssymb,
 aps,
 prd
]{revtex4-2}

\usepackage{graphicx}% Include figure files
\usepackage{dcolumn}% Align table columns on decimal point
\usepackage{bm}% bold math
\usepackage{lipsum}

\newcommand{\rcite}[1]{Ref.~\cite{#1}}
\newcommand{\eq}[1]{Eq.~(\ref{#1})}
\newcommand{\elve}{{E_{\rm LV,E}}}
\newcommand{\elvi}{{E_{\rm LV,I}}}
\newcommand{\elvc}{{E_{\rm LV,C}}}
\newcommand{\ev}{{~\rm eV}}
\newcommand{\gev}{{~\rm GeV}}
\newcommand{\pev}{{~\rm PeV}}

\begin{document}

\preprint{APS/123-QED}

\title{Exploring Lorentz Violation in Spacetime through Universal Finsler Geometry
%Considerations of Lorentz Violation from Flat Finsler Spacetime
}% Force line breaks with \\
% \thanks{A footnote to the article title}%

\author{Jie Zhu}
 \email{jiezhu@cqu.edu.cn}
 % \altaffiliation[Also at ]{Physics Department, XYZ University.}%Lines break automatically or can be forced with \\
\author{Hao Li}%
 \email[Corresponding author: ]{haolee@cqu.edu.cn}
\affiliation{%
 % Authors' institution and/or address\\
 % This line break forced with \textbackslash\textbackslash
Department of Physics and Chongqing Key Laboratory for Strongly Coupled Physics, Chongqing University, Chongqing 401331, P.R. China
}%

% \collaboration{MUSO Collaboration}%\noaffiliation
\author{Bo-Qiang Ma}
\email[Corresponding author: ]{mabq@pku.edu.cn}
\affiliation{School of Physics,
    %and State Key Laboratory of Nuclear Physics and Technology, 
    Peking University, Beijing 100871, China}
\affiliation{School of Physics, Zhengzhou University, Zhengzhou 450001, China}

% \affiliation{Center for High Energy Physics, Peking University, Beijing 100871, China}
% \affiliation{School of Physics, Peking University}
% \author{Charlie Author}
%  \homepage{http://www.Second.institution.edu/~Charlie.Author}
% \affiliation{
%  Second institution and/or address\\
%  This line break forced% with \\
% }%
% \affiliation{
%  Third institution, the second for Charlie Author
% }%
% \author{Delta Author}
% \affiliation{%
%  Authors' institution and/or address\\
%  This line break forced with \textbackslash\textbackslash
% }%

% \collaboration{CLEO Collaboration}%\noaffiliation

% \date{\today}% It is always \today, today,
             %  but any date may be explicitly specified

\begin{abstract}

Finsler geometry serves as a fundamental and natural extension of Riemannian geometry, providing a valuable framework for investigating Lorentz violation in spacetime. Previous studies have treated the Finsler structures associated with different particles as distinct entities. In this paper, we propose a novel hypothesis suggesting that the Finsler structure may represent an intrinsic property of the universe itself. Under this assumption, we derive a series of modified dispersion relations that have not been previously explored, and we analyze their implications. Our findings indicate that the scales of Lorentz violation for massive particles are proportional to their masses. Furthermore, we demonstrate that this hypothesis aligns well with existing phenomenological results regarding Lorentz violation observed in photons, neutrinos, and electrons.

%Finsler geometry is a natural and fundamental generalization of Riemann geometry, and is a tool to research Lorentz violation. 
%In previous research, the Finsler structures of different particles are considered to be different.
%Based on several considerations, we propose a hypothesis that the Finsler structure may be the property of the universe.
%In this assumption, we find a series of modified dispersion relations that have never been studied before, and we study their properties.
%We find that the Lorentz violation scales for massive particles are proportional to their masses.
%We also find that this hypothesis is consistent with the current results of Lorentz violation from photons, neutrinos, and electrons.

% Lorentz Violation\\
% Finsler\\
% Consider LV in Finsler\\
% Reach the result that
% An article usually includes an abstract, a concise summary of the work
% covered at length in the main body of the article. 
% \begin{description}
% \item[Usage]
% Secondary publications and information retrieval purposes.
% \item[Structure]
% You may use the \texttt{description} environment to structure your abstract;
% use the optional argument of the \verb+\item+ command to give the category of each item. 
% \end{description}
\end{abstract}

%\keywords{Suggested keywords}%Use showkeys class option if keyword
                              %display desired

\maketitle

%\tableofcontents

\section{Introduction}\label{sec:1}

%Lorentz invariance is a basic assumption of Einstein's relativity, and it is the foundation of general relativity and quantum field theory in modern physics.
%However, in quantum gravity (QG)~\cite{Amelino-Camelia:2004qyt}, Lorentz violation (LV) may happen near the Planck scale ($E_{\rm Pl}\simeq 1.22\times 10^{19}\gev$).
%There are many models and theories including LV effects,  including the QG theories such as string theory~\cite{Danielsson:2001et, Li:2021gah,Li:2022sgs,Li:2023wlo} and loop quantum gravity (LQG)~\cite{Rovelli:2008zza, Ashtekar:2021kfp, Li:2022szn}, spacetime structure theories such as doubly special relativity (DSR)~\cite{Amelino-Camelia:2000cpa, Amelino-Camelia:2000stu, Magueijo:2001cr, Magueijo:2002am} and very special relativity (VSR)~\cite{Cohen:2006ky},  and the effective theory with extra terms, such as the standard-model extension (SME)~\cite{Colladay:1996iz, Colladay:1998fq, Kostelecky:2003fs}.
%For a recent review on Lorentz invariance violation, see~\cite{He:2022gyk}.

Lorentz invariance is a fundamental assumption of Einstein's theory of relativity and serves as the cornerstone of both general relativity and quantum field theory in modern physics. However, in the context of quantum gravity (QG)~\cite{Amelino-Camelia:2004qyt}, Lorentz violation (LV) may occur near the Planck scale ($E_{\rm Pl}\simeq 1.22\times 10^{19}\gev$). Numerous models and theories that incorporate LV effects have been proposed, including QG frameworks such as string theory~\cite{Danielsson:2001et, Li:2021gah, Li:2022sgs, Li:2023wlo} and loop quantum gravity (LQG)~\cite{Rovelli:2008zza, Ashtekar:2021kfp, Li:2022szn}, as well as theories concerning spacetime structure, such as doubly special relativity (DSR)~\cite{Amelino-Camelia:2000cpa, Amelino-Camelia:2000stu, Magueijo:2001cr, Magueijo:2002am} and very special relativity (VSR)~\cite{Cohen:2006ky}. Additionally, effective theories with extra terms, such as the standard model extension (SME)~\cite{Colladay:1996iz, Colladay:1998fq, Kostelecky:2003fs}, have been developed to account for these effects. For a comprehensive review of Lorentz invariance violation, see~\cite{He:2022gyk}.

%A key feature of LV is that particles admit modified dispersion relations (MDRs) like
A key feature of LV is the existence of modified dispersion relations (MDRs) for particles, which can be expressed as follows:
\begin{equation}
    \begin{aligned}
E^2 & =m^2+p^2+D(p, \mu, M) \\
& =m^2+p^2+\sum_{n=1}^{\infty} \alpha_n(\mu, M) p^n,
\end{aligned}
\end{equation}
%where $\alpha_n$ are dimensional coefficients, $\mu$ is some particle physics mass scale, and $M$ is the scale associated with the new physics responsible for the correction of the dispersion relation.
where $\alpha_n$ are dimensional coefficients, 
$\mu$ is a particle physics mass scale, and 
$M$ is the scale associated with the new physics responsible for the modifications to the dispersion relation.

%The spacetime of gravity is described by general relativity (GR) in the language of Riemann geometry.
%However, since GR is a Lorentz invariant theory, it cannot describe the spacetime of Lorentz violation.
%The early temptation to describe the spacetime of LV is using a rainbow metric~\cite{Magueijo:2002xx}, which depends on the energy of the particle. 
%This is a natural concept from the QG point of view and arose in different contexts such as spacetime foam~\cite{Ellis:2008gg}, the renormalization group applied to gravity~\cite{Girelli:2006sc},  or as a consequence of averaging over QG fluctuations~\cite{Aloisio:2005qt}.
%However, the rainbow metric still lacks a rigorous formulation since it involves a metric defined on the tangent bundle while depending on a quantity associated with the cotangent bundle (the 4-momentum of the particle).

In general relativity (GR), the fabric of spacetime is described using Riemann geometry. However, because GR is a Lorentz-invariant theory, it cannot adequately capture the characteristics of spacetime that exhibit Lorentz violation. An early approach to describe LV spacetime involved the concept of a rainbow metric~\cite{Magueijo:2002xx}, which varies with the energy of the particle. This idea arises naturally from the perspective of QG and has been explored in various contexts, such as spacetime foam~\cite{Ellis:2008gg}, the renormalization group applied to gravity~\cite{Girelli:2006sc}, and as a result of averaging over QG fluctuations~\cite{Aloisio:2005qt}. Nevertheless, the rainbow metric lacks a rigorous formulation, as it involves a metric defined on the tangent bundle while depending on a quantity associated with the cotangent bundle (the 4-momentum of the particle).

In the quest for geometrical structures that lead to MDRs, researchers have moved beyond Riemannian geometry to adopt a generalization known as Finsler geometry~\cite{Girelli:2006fw}. The connections between Finsler geometries and LV theories have been well established in the literature, including applications in SME~\cite{Kostelecky:2011qz, AlanKostelecky:2012yjr, Colladay:2015wra, Russell:2015gwa, Schreck:2015seb, Edwards:2018lsn}, VSR~\cite{Gibbons:2007iu}, and DSR~\cite{Amelino-Camelia:2014rga}. Finsler geometry serves as an effective framework for describing LV spacetime and is particularly useful for studying particles in gravitational backgrounds, such as in the time-delay formula for astroparticles~\cite{Zhu:2022blp, Zhu:2023mps}. A recent review of Finsler geometry's applications in LV can be found in Ref.~\cite{Zhu:2023kjx}. Additionally, Finsler geometry has potential applications in other areas of new physics, including Finsler gravity~\cite{Li:2010zc, Pfeifer:2011xi}, dark matter~\cite{Chang:2008yv}, dark energy~\cite{Basilakos:2013hua, Chang:2009pa}, and bounce cosmology~\cite{Minas:2019urp}. Thus, Finsler geometry merits further exploration in the context of new physics.

Unlike Riemann geometry, which defines an inner product structure over the tangent bundle, Finsler geometry is based on the so-called Finsler structure or Finsler norm, denoted as $F$, satisfying the property $F(x,\lambda y)=\lambda F(x,y)$ for all $\lambda>0$, where $x \in M$  represents the position and $y \equiv \frac{d x}{d \tau}$ represents the velocity. The Finsler metric is defined as follows:
\begin{equation}
g_{\mu \nu} \equiv \frac{\partial}{\partial y^{\mu}} \frac{\partial}{\partial y^{\nu}}\left(\frac{1}{2} F^{2}\right).\label{eq:gmn}
\end{equation}
Finsler geometry originates from integrals of the form:
\begin{equation}
    \int_{a}^{b} F\left(x^{1}, \cdots, x^{n} ; \frac{d x^{1}}{d \tau}, \cdots, \frac{d x^{n}}{d \tau}\right) d \tau. \label{eq:F}
\end{equation} 
The Finsler structure represents the length element of Finsler space. If $F^2$ is quadratic in 
$y$, the Finsler metric $g_{\mu \nu}$ becomes independent of $y$, and the Finsler geometry effectively reduces to Riemann geometry, at which point the Finsler metric is referred to as Riemannian. To describe the ``1+3" spacetime, we transition from Finsler geometry to pseudo-Finsler geometry. In the following sections, we will refer to both pseudo-Finsler geometry and Finsler geometry collectively as Finsler geometry.

\section{Finsler Spacetime from Lorentz Violation}\label{sec:2}

% Finsler geometry can describe the spacetime in which particles with Lorentz violation live and 
Plenty of literature has studied the Finsler norm of a given MDR with Lorentz violation  ~\cite{Girelli:2006fw, Amelino-Camelia:2014rga, Lobo:2016xzq, Lobo:2016lxm, Lobo:2020qoa, Zhu:2022blp, Zhu:2023mps}. 
The method to obtain the Finsler norm corresponding to an MDR $\mathcal{M}(p)=m^2$ is described in \rcite{Girelli:2006fw} in detail, and the procedure is shown as follows.
First, write the action of the particle as
\begin{equation}
    I=\int\left(\dot{x}^\mu p_\mu-\lambda(\mathcal{M}(p)-m^2)\right) d\tau.\label{eq:original_action}
\end{equation}
Then, we can obtain the relation that
\begin{equation}
    \dot{x}^\mu=\lambda\frac{\partial\mathcal{M}}{\partial p_\mu}.\label{eq:xp}
\end{equation}
We can rewrite the above expression to express $p$ as the function of $\dot{x}$ and $\lambda$, and then the action~(\ref{eq:original_action}) can be expressed as
\begin{equation}
    I=\int\mathcal{L}(\dot{x},\lambda)d\tau.
\end{equation}
Varying the action with respect to the parameter $\lambda$, we can get $\lambda$ as the function of $\dot{x}$ as $\lambda=\lambda(\dot{x})$, thus
 \begin{equation}
    I=\int\mathcal{L}(\dot{x},\lambda(\dot{x}))d\tau.
\end{equation}
Finally, from the relation
\begin{equation}
    I=m \int F(x,\dot{x})d\tau,
\end{equation}
we obtain the Finsler norm as
\begin{equation}
    mF(x,y)=\mathcal{L}(y,\lambda(y)).
\end{equation}
Here, we can easily see that the relation $v=\frac{dx}{dt}=\frac{|\dot{\vec{x}}|}{\dot{x^0}}=\frac{\partial E}{\partial p}$ holds from \eq{eq:xp}.

For most cases of MDRs, it is hard to express the corresponding Finsler norms explicitly. In experiments, people only consider the leading correction to dispersion relations, so the above calculation is carried out perturbatively. 
For a general MDR expressed in leading order as
\begin{equation}
p_0^2=m^2+\vec{p}~^2+\alpha h(p_0,\vec{p}),\label{eq:MDR_general}
\end{equation}
where $h(p_0,\vec{p})$ is an order $n+2$ homogeneous function of $p^\mu$, i.e., $h(\lambda p_0,\lambda\vec{p})=\lambda^{n+2} h(p_0,\vec{p})$, and the mass dimension for the parameter $\alpha$ is $[\alpha]=-n$. 
Here, $n$ represents the broken order of LV.
The corresponding Finsler norm is derived in \rcite{Zhu:2023mps} by a different method in leading order as
\begin{equation}
    F=\sqrt{\mathcal{Y}^2}+\frac{\alpha m^n}{2}\frac{h(y^0,-\vec{y})}{(\mathcal{Y}^2)^{\frac{n+1}2}},\label{eq:Finsler_general}
\end{equation}
where $\mathcal{Y}^2=\eta_{\mu\nu}y^\mu y^\nu=(y^0)^2-\vec{y}~^2$.
For example, the MDR of DSR-1~\cite{Amelino-Camelia:2000cpa, Amelino-Camelia:2000stu} can be expressed in leading order as
\begin{equation}
    p_0^2=m^2+\vec{p}~^2+\ell p_0\vec{p}~^2,
\end{equation}
here $h(p_0,\vec{p})=p_0\vec{p}~^2$ and $n=1$,
and thus, the Finsler norm is
\begin{equation}
    F=\sqrt{\mathcal{Y}^2}+\frac{\ell m}{2}\frac{y_0 \vec{y}~^2}{\mathcal{Y}^2},
\end{equation}
which is exactly the result in \rcite{Amelino-Camelia:2014rga}.

%As discussed in \rcite{Girelli:2006fw}, even assuming a constant $\alpha$ for different particles in \eq{eq:MDR_general} still the corresponding Finsler norm \eq{eq:Finsler_general} is different.
%This is because the Finsler norm has no mass scales, and thus, the mass dimension of parameter $\alpha$ should be canceled by another massive parameter, and this parameter is the mass of the particle. 
%This also means that Finsler geometry can hardly describe the MDR for a massless particle with a parameter with dimensions of mass, for there exists no other parameter to cancel the mass dimension in the Finsler norm.
%In this work, we mainly consider massive particles. 
%If we have to deal with massless particles, we obtain the result of massive particles first and take the limit as the mass approaches zero.

As discussed in \rcite{Girelli:2006fw}, even if we assume a constant $\alpha$ for different particles in \eq{eq:MDR_general}, the corresponding Finsler norm in \eq{eq:Finsler_general} remains distinct. This discrepancy arises because the Finsler norm lacks intrinsic mass scales; therefore, the mass dimension of the parameter $\alpha$ must be balanced by another massive parameter, specifically the mass of the particle. Consequently, Finsler geometry is not well-suited to describe the modified dispersion relation (MDR) for massless particles when using a parameter with dimensions of mass, as there is no other parameter available to cancel the mass dimension within the Finsler norm. In this work, we primarily focus on massive particles. If we need to address massless particles, we first derive the results for massive particles and then take the limit as the mass approaches zero.

%One more comment concerns that \eq{eq:MDR_general} and \eq{eq:Finsler_general} could be the leading order of the asymptotic series rather than the Taylor series of the full MDR and corresponding Finsler norm, which means that if we solve the Finsler norm in higher order, the result may deviate further from the right result. In this work, we only consider the contribution of the leading order.

Additionally, it is important to note that \eq{eq:MDR_general} and \eq{eq:Finsler_general} may represent the leading order of an asymptotic series rather than the Taylor series expansion of the full MDR and corresponding Finsler norm. This implies that if we were to consider higher-order terms in the Finsler norm, the results could deviate further from the accurate outcome. In this study, we restrict our attention to the contributions from the leading order only.

\section{Lorentz Violation from Finsler Spacetime}\label{sec:3}
%Now, let's consider the correspondence of \eq{eq:MDR_general} and \eq{eq:Finsler_general} from a different perspective.
%It is really strange that different particles feel different Finsler structures. 
%So, what if the Finsler structure is the same for all of the particles, i.e., the Finsler structure is the property of the universe?
%Here, we propose a hypothesis: 
%\begin{itemize}
%  \item If there is Lorentz violation, the spacetime of Lorentz violation can be effectively described by a unique Finsler geometry. The leading order of the Finsler norm can be expressed as \eq{eq:Finsler_general}.
%\end{itemize}

Now, let us examine the correspondence between \eq{eq:MDR_general} and \eq{eq:Finsler_general} from a different perspective. It is indeed peculiar that different particles experience distinct Finsler structures. This raises the question: what if the Finsler structure is uniform across all particles, implying that it is an intrinsic property of the universe?

To explore this idea, we propose the following hypothesis:

\begin{itemize} \item If Lorentz violation exists, the spacetime associated with this violation can be effectively described by a unique Finsler geometry. The leading order of the Finsler norm can be expressed as \eq{eq:Finsler_general}. \end{itemize}

%The above hypothesis comes from the following considerations. 
%The first one is from cosmology.
%Finsler geometry has shown its strength in solving cosmology problems, and it also describes the spacetime of Lorentz violation. 
%One natural idea is to construct a Finsler geometry that simultaneously describes Lorentz violation and cosmology, and thus, there should be only one Finsler norm in the Universe. 
%In flat spacetime, the full Finsler norm only contains Lorentz violation.
%Suppose there is one kind of particle with Lorentz violation, and its MDR can be expressed in leading order as \eq{eq:MDR_general}, then the Finsler norm in leading order is \eq{eq:Finsler_general}.
%Since there is only one Finsler norm, then the Finsler norm for the flat Universe should be \eq{eq:Finsler_general}.

This hypothesis is motivated by several considerations. The first stems from cosmology. Finsler geometry has demonstrated its utility in addressing cosmological problems and is capable of describing the spacetime associated with Lorentz violation. A natural approach is to construct a Finsler geometry that encompasses both Lorentz violation and cosmological dynamics, suggesting that there should be a single Finsler norm governing the universe.

In flat spacetime, the complete Finsler norm encompasses only Lorentz violation. Suppose we consider a specific type of particle that experiences Lorentz violation, with its modified dispersion relation (MDR) expressible in leading order as \eq{eq:MDR_general}. In this case, the leading order Finsler norm would be represented by \eq{eq:Finsler_general}. Given that there is only one Finsler norm, it follows that the Finsler norm for a flat universe should indeed be \eq{eq:Finsler_general}.

Another consideration comes from the speed of a group of particles.
Suppose there are $N$ relatively stationary particles of the same kind moving at speed $v$ relative to the laboratory.
Suppose the MDR of the particles can be expressed as
\begin{equation}
    E^2=m^2+p^2+\alpha h(E,p),
\end{equation}
and for the group, the MDR is
\begin{equation}
    E^2=m^2+p^2+\alpha' h'(E,p),
\end{equation}
where $h$ and $h'$ are order $n+2$ and $n'+2$ homogeneous functions.
As mentioned in Sec.~\ref{sec:2}, the relation $v=\frac{\partial E}{\partial p}$ holds in Finsler geometry, we get
\begin{equation}
    v=\frac{2p+\alpha h_2(E,p)}{2E-\alpha h_1(E,p)},\label{eq:speed_single}
\end{equation}
where $h_1(E,p)=\frac{\partial}{\partial E}h(E,p)$ and $h_2(E,p)=\frac{\partial}{\partial p}h(E,p)$,
and $h_1(E,p)$ and $h_2(E,p)$ are order $n+1$ homogeneous functions.
The composition law of the 4-momentum in Finsler geometry may be modified~\cite{Lobo:2022wrx}, 
but if the speed $v$ is not too fast, we can still use the classical adding law, i.e., the total energy of the group is $E'=NE$, the total mass is $m'=Nm$, and the total momentum is $p'=Np$. So the speed of the group of particles is
\begin{eqnarray*}
v&&=\frac{2p'+\alpha' h'_2(E',p')}{2E'-\alpha' h'_1(E',p')}\\
&&=\frac{2Np+\alpha' h'_2(NE,Np)}{2NE-\alpha' h'_1(NE,Np)}\\
&&=\frac{2p+\alpha'N^{n'} h'_2(E,p)}{2E-\alpha'N^{n'} h'_1(E,p)}.
\end{eqnarray*}
The speed of the group should be equal to the speed of a single particle,
thus we have
\begin{equation}
    \alpha'N^{n'}h'(E,p)=\alpha h(E,p).
\end{equation}
So $h'(E,p)$ is proportional to $h(E,p)$.
Letting $h'(E,p)=h(E,p)$, we have
\begin{equation}
    n=n', \quad \alpha'=\alpha N^{-n}.
\end{equation}
Now we can see that $\alpha' m'^{n'}=\alpha m^n$, and the Finsler norm of the group of particles is exactly the same as one particle.

The last consideration comes from the modified conservation law of DSR.
For example, in DSR-2, the modified conservation law is suggested by Magueijo and Smolin as~\cite{Magueijo:2002am}
\begin{equation}
    \pi^\mu_{\rm total}=\sum_i \frac{p_\mu^{(i)}}{1-\lambda p_0^{(i)}},
\end{equation}
where $p_\mu$ is 4-momentum and the index $i$ is for particles.
This conservation law meets challenges when applied to macroscopic objects,
because $\lambda=1/E_{\rm DSR} \sim 1/E_{\rm Pl}$ and the mass of macroscopic objects can easily reach the Plank scale. 
To solve this problem, Magueijo and Smolin suggest that for an $N$-particle system, 
the DSR parameter $\lambda$ should be replaced by $\lambda/N$,
and the total 4-momentum for the group satisfies
\begin{equation}
    \frac{p_\mu^{(N)}}{1-\frac{\lambda p_0^{(N)}}N}=\sum_i \frac{p_\mu^{(i)}}{1-\lambda p_0^{(i)}}.
\end{equation}
In fact, replacing $\lambda$ with $\lambda/N$ is the special case of $\alpha'=\alpha N^{-n}$.
Again, the modified conservation law suggested that the Finsler norm should be the same for a single particle and a group of particles.

Though the considerations only contain one kind of particle, 
we can step further to assume that is true for all kinds of particles, or the Finsler norm is the property of the universe.
Suppose the spatial rotational symmetry holds, we rewrite \eq{eq:Finsler_general} the leading order of the Finsler norm as
\begin{equation}
    F=\sqrt{\mathcal{Y}^2}+\frac{A}{2}\frac{h(y^0,|\vec{y}|)}{(\mathcal{Y}^2)^{\frac{n+1}{2}}},\label{eq:Finsler_main}
\end{equation}
where $A$ is a constant of the Universe without mass dimension, and $h(y^0,\vec{y})$ is an order $n+2$ homogeneous function, $n>0$.
For a Finsler norm expressed as \eq{eq:Finsler_main},
we can assume that $h(1,1)=1$.
If $h(1,1)\neq 0$, we can absorb $h(1,1)$ into parameter $A$ and define a new $h'(y^0,|\vec{y}|)=h(y^0,|\vec{y}|)/h(1,1)$.
If $h(1,1)=0$, assume that equation $h(y^0,|\vec{y}|)=0$ has a root $y^0=|\vec{y}|$ of multiplicity $l$,
we can define a new $h'(y^0,|\vec{y}|)$ as
\begin{equation}
    h'(y^0,|\vec{y}|)=\frac{h(y^0,|\vec{y}|)}{(\mathcal{Y}^2)^l},
\end{equation}
and replace the $n$ in \eq{eq:Finsler_main} by $n-2l$.
So we can always assume that $h(1,1)=1$ in \eq{eq:Finsler_main}.
Under the condition $h(1,1)=1$, the sign of $A$ reflects whether the LV is subluminal or superluminal.
If $A<0$, then all particles are subluminal, and if $A>0$, the speed of particles could be superluminal.
In the picture that the Finsler norm is the property of the universe, the Lorentz violation of all particles is either entirely subluminal or superluminal.
Also, in \eq{eq:Finsler_main}, $|A|\ll 1$, and the modified term matters when $\mathcal{Y}^2/(y^0)^2\sim 0$, which corresponds to $v\sim 1$ and high energy of particles.

Now we derive the MDR from the Finsler norm \eq{eq:Finsler_main}.
% Suppose $h(1,1)\neq 0$.
Directly, one solution is
\begin{equation}
    E^2=m^2+p^2+Am^{-n} h(E,p).
\end{equation}
However, notice that \eq{eq:Finsler_main} can be expressed as
\begin{equation}
    F=\sqrt{\mathcal{Y}^2}+\frac{A}{2}\frac{h(y^0,|\vec{y}|)(\mathcal{Y}^2)^k}{(\mathcal{Y}^2)^{\frac{n+1}{2}+k}},\label{eq:sol1}
\end{equation}
we found a series of solutions as
\begin{equation}
    E^2=m^2+p^2+Am^{-n-2k} h(E,p)(E^2-p^2)^k.\label{eq:sol2}
\end{equation}
This solution reduces to \eq{eq:sol1} at low energies, 
since $Am^{-n-2k}(E^2-p^2)^k=Am^{-n-2k}(m^2)^k+o(A)=Am^{-n}+o(A)$.
But at high energies, these MDRs behave differently.
$k$ must satisfy $k\geq-\frac{n}{2}$,
because we want the modified term in MDR to be suppressed by a mass scale.
When $k=-\frac{n}{2}$, \eq{eq:sol2} becomes
\begin{equation}
    E^2=m^2+p^2+A\frac{h(E,p)}{(E^2-p^2)^{\frac{n}{2}}}.
\end{equation}
In this special case, no energy scale except the mass $m$ appears.
% From \eq{eq:speed_single}, we can see that if $A<0$, 

\section{Physical Consequences}\label{sec:4}

\subsection{Time Delay}\label{sec:4.1}

One method to test LV is by the arrival time differences from high-energy astroparticles. 
In our previous work~\cite{Zhu:2022blp, Zhu:2023mps}, 
we derive the arrival time difference between a high-energy particle and a low-energy one.
For the Finsler norm as \eq{eq:Finsler_main} with $h(1,1)=1$,
the arrival time delay is
\begin{equation}
    \Delta t=A \left(\frac{E_\mathrm{obs}}{m}\right)^n\frac{n+1}{2H_0}\int_0^z\frac{(1+z^{\prime})^n}{\sqrt{\Omega_m\left(1+z^{\prime}\right)^3+\Omega_\Lambda}}dz^{\prime},
\end{equation}
where $z$ is the redshift of the source of the two particles, $E_\mathrm{obs}$ is the observed energy of the high-energy particle from Earth equipment, $\Omega_{\rm{m}}$ and $\Omega_{\rm{\Lambda}}$ are universe constants, and $H_0$ is the current Hubble parameter.
In Ref.~\cite{Zhu:2023mps}, we state that if $h(1,1)=0$, 
there is no arrival time delay.
But here, we find we can still absorb the roots of $h(y^0,|\vec{y}|)$ into the denominator of the modified term in \eq{eq:Finsler_main}. 
Here we correct the result as that the situation where the arrival time difference disappears is when the modified term of \eq{eq:Finsler_main} does not have the pole $y^0=|\vec{y}|$.
For DSR-2, the Finsler norm is of Randers type and the modified term does not have the pole $y^0=|\vec{y}|$, so there is no time delay.

We can see that the time delay formula only depends on the homogeneous order of $h(y^\mu)$. 
So, for a fixed Finsler norm, no matter how the MDR varies, the time delay formula is the same and only depends on $n$.
Here, we call $n$ the \textit{implicit violation order of Lorentz invariance} as $n_{\rm I}$.
Correspondingly, the \textit{explicit violation order of Lorentz invariance} $n_{\rm E}$ is read off the MDR. 
For example, the explicit violation order of \eq{eq:sol2} is $n_{\rm E}=n+2k$ for the modified term $h(E,p)(E^2-p^2)^k$ has mass dimension $n+2k$.
Then, we can define the implicit LV scale $\elvi$ and the explicit LV scale $\elve$ as
% \begin{equation}
%     \elvi=m |A|^{-1/n_{\rm I}}, \quad \elve=m |A|^{-1/n_{\rm E}},\label{eq:elv}
% \end{equation}
% with $n_{\rm I}=n$ and $n_{\rm E}=n+2k$.
\begin{equation}
    \elvi=m |A|^{-1/n}, \quad \elve=m |A|^{-1/(n+2k)}.\label{eq:elv}
\end{equation}
In this notation, the time delay formula is
\begin{equation}
    \Delta t= s\left(\frac{E_\mathrm{obs}}{\elvi}\right)^n\frac{n+1}{2H_0}\int_0^z\frac{(1+z^{\prime})^n}{\sqrt{\Omega_m\left(1+z^{\prime}\right)^3+\Omega_\Lambda}}dz^{\prime},\label{eq:dt}
\end{equation}
where $s=\mathrm{sign}(A)$,
and the MDR is
\begin{equation}
    E^2=m^2+p^2+s\frac{1}{\elve^{n+2k}}h(E,p)(E^2-p^2)^k.\label{eq:mdr}
\end{equation}
$\elvi$ and $n_{\rm I}$ describe the energy scale and the broken order of LV that can be tested from the time delay of astroparticles, and for a fixed Finsler norm $n_{\rm I}$ is the same for all kinds of particles.
$\elve$ and $n_{\rm E}$ describe the energy scale and the broken order of LV from the MDR.
% , and we will see that it will affect the test of threshold anomalies. 
% In previous research, people think that the energy scales of LV tested from time delays and threshold anomalies are the same,
% but now we will see that they could be different.
If $-n/2<k<0$, then $\elve>\elvi$, and if $k>0$, then $\elve<\elvi$.
Only when $k=0$ are the two scales the same.
$\elve$ and $\elvi$ have the relation
\begin{equation}
    \frac{\elve}{\elvi}=|A|^\frac{2k}{n(n+2k)}.\label{eq:eei}
\end{equation}
Since Eqs.~(\ref{eq:dt}), (\ref{eq:mdr}) and (\ref{eq:eei}) are free of $m$, we may assume that the above relations hold for massless particles like photons.

% Another thing that we should notice is that f
%From \eq{eq:elv}, we can see that for massive particles, the LV scales are proportional to their masses! 
%That means the greater the mass, the higher the scales of LV effects, and consequently, the more difficult it becomes to test for them.

From \eq{eq:elv}, we observe that for massive particles, the Lorentz violation (LV) scales are proportional to their masses. This implies that as the mass increases, the scales of LV effects also rise, making it increasingly challenging to test for these effects experimentally.

\subsection{Behavior of MDRs}\label{sec:4.2}

In this part, we discuss the behavior of the MDRs of \eq{eq:mdr} that come from the Finsler norm \eq{eq:Finsler_main}.
We are concerned about how the MDRs behave in the high-energy region, i.e., the asymptotic expansion of $E^2-p^2$ as a function of $E$,
because $E^2-p^2$ plays an important role in threshold anomaly calculations. 
In the high-energy region, $E$ is close to $p$, and $h(E,p)=E^{n+2}h(1,p/E)\simeq E^{n+2}h(1,1)+O(A)=E^{n+2}+O(A)$,
and the MDRs are reduced to
\begin{equation}
    E^2=m^2+p^2+s\frac{1}{E_{\rm LV,E}^{n+2k}}E^{n+2}(E^2-p^2)^k.\label{eq:mdrE}
\end{equation}
The above MDR is considered to be effective only when $E\ll \elve$.
We can asymptotic solve \eq{eq:mdrE} as
\begin{widetext}
\begin{eqnarray}
E^2-p^2\simeq&&m^2+Am^{-n}E^{n+2}+kA^2m^{-2n-2}E^{2n+4}+O(A^3)\nonumber\\
  =&&m^2+ E^2\left(s\left(\frac{E}{\elvi}\right)^n+k\left(\frac{E}{\elvi}\right)^{2n-\frac{n}{k}}\left(\frac{E}{\elve}\right)^{2+\frac{n}{k}} \right)+O(A^3).\label{eq:MDR_expand}
\end{eqnarray}
\end{widetext}
It seems that the ordinary relation 
\begin{equation}
E^2-p^2\simeq m^2 +sE^2 \left(\frac{E}{\elvi}\right)^n\label{eq:mdr_ord}   
\end{equation}
holds from the above solution.
However, from the above asymptotic series, we can get the scale below which the ordinary relation holds.
An asymptotic series fails when the next-to-leading order is comparable to the leading order.
Let the first term equal to the second term or let the second term equal to the third term, 
if $|k|\sim O(1)$,
we know that the ordinary relation \eq{eq:mdr_ord} fails when
$|A|m^{n+2}E^{n+2}\sim 1$, or
\begin{equation}
    E\sim \elvc=m|A|^{-\frac{1}{n+2}},
\end{equation}
where ``C'' denotes ``critical''.
We can see that $\elvc \ll \elvi$ for
% \begin{equation}
%     \frac{\elvc}{\elvi}=|A|^\frac{1}{n(n+2)}
% \end{equation}
${\elvc}/{\elvi}=|A|^\frac{1}{n(n+2)}$,
$|A|\ll 1$, and $n>0$.
So we can see that if $|k|\sim O(1)$, then \eq{eq:mdr_ord} holds only below a scale much smaller than $\elvi$.
When $|k|\ll 1$, the scale where \eq{eq:mdr_ord} holds moves to $|k|^{-\frac{1}{n+2}}\elvc$.
If $k=0$, the MDR \eq{eq:mdrE} reduces to \eq{eq:mdr_ord}, 
and in this case, the scale is thought to be $\elvi$.
From the above discussion, we see that if $k\neq 0$, the behavior of $E^2-p^2$ is quite different from the ordinary relation \eq{eq:mdr_ord} above the scale $\elvc$.
% and we should deal with different $k$ differently.

We first consider the superluminal case, i.e., $A>0$.
% In the superluminal case $A>0$, 
If $E\gg\elvc$, then $E^2-p^2\gg m^2$, thus
\begin{equation}
    E^2-p^2\simeq Am^{-n-2k} E^{n+2}(E^2-p^2)^k,
\end{equation}
hence
\begin{equation}
    E^2-p^2\simeq A^\frac{1}{1-k}m^{2-\frac{n+2}{1-k}}E^\frac{n+2}{1-k}=m^2 \left(\frac{E}{\elvc}\right)^\frac{n+2}{1-k}.\label{eq:sol_over}
\end{equation}
This solution is effective only for $k<1$.
Obviously, $E^2-p^2\gg m^2$ if $E\gg\elvc$.
Combining the expansion \eq{eq:MDR_expand}, we see that as the energy gets larger, $E^2-p^2$ gets larger.
% An example is the superluminal case $A>0$ and $k=-n/2$.
A special case is $k=-n/2$.
For energy $E\gg \elvc$, the asymptotic series is
\begin{equation}
    E^2-p^2=A^\frac{2}{n+2}E^2+o(A),
\end{equation}
and also $E^2-p^2\gg A^\frac{2}{n+2} \elvc^2=m^2$.

% An interesting case in
For the subluminal case, if $k<0$,
Denote $E^2-p^2$ as $S$,
we have
\begin{eqnarray}
m^2&&=S+|A|m^{-n+2|k|}E^{n+2}S^{-|k|}\label{eq:S}\\
&&\geq (|k|+1)\left[\left(\frac{S}{|k|}\right)^{|k|}|A|m^{-n+2|k|}E^{n+2}S^{-|k|}\right]^\frac{1}{|k|+1}\nonumber\\
&&=(|k|+1)\left[{|k|}^{-|k|}|A|m^{-n+2|k|}E^{n+2}\right]^\frac{1}{|k|+1}\nonumber,
\end{eqnarray}
and thus
\begin{equation}
    E\leq C(k) m |A|^\frac{2}{n+2}=C(k)\elvc,
\end{equation}
where $C(k)=|k|^\frac{|k|}{n+2}(|k|+1)^{-\frac{|k|+1}{n+2}}\leq 1$
and $C(k)\geq \frac{2^{\frac{1}{n+2}} n^{\frac{n}{2 n+4}}}{\sqrt{n+2}}\geq \frac{1}{\sqrt{2}}$ for $-n/2\leq k< 0$.
In this case, the energy of the particles is limited below $\elvc$.
Besides, for a given $p$, there are two solutions for $E$.
If the main contribution for $m^2$ in \eq{eq:S} comes from $E^2-p^2$,
then the expansion is \eq{eq:mdr_ord}.
Or, if the main contribution for $m^2$ in \eq{eq:S} comes from the other term, then we have
\begin{equation}
    m^2\simeq |A|m^{-n+2|k|}E^{n+2}(E^2-p^2)^{-|k|}.
\end{equation}
In this case
% \begin{eqnarray}
%     E^2-p^2&&\simeq |A|^\frac{1}{|k|}m^{2-\frac{n+2}{|k|}}E^\frac{n+2}{|k|}\nonumber\\
%     &&\ll |A|^\frac{1}{|k|}m^{2-\frac{n+2}{|k|}}\elvc^\frac{n+2}{|k|}\nonumber\\
%     &&=m^2\nonumber.
% \end{eqnarray}
\begin{equation}
    E^2-p^2\simeq |A|^\frac{1}{|k|}m^{2-\frac{n+2}{|k|}}E^\frac{n+2}{|k|}=m^2\left(\frac{E}{\elvc}\right)^\frac{n+2}{|k|}.\label{eq:sol_unp}
\end{equation}
Thus if $E\ll\elvc$, then $E^2-p^2\ll m^2$.
% \begin{equation}
%     E^2-p^2\ll |A|^\frac{1}{|k|}m^{2-\frac{n+2}{|k|}}\elvc^\frac{n+2}{|k|}=m^2.
% \end{equation}
It is hard to imagine that $E^2-p^2=m^2$ breaks in the low-energy region, so this solution may be unphysical.

If $k>0$ and $E\gg\elvc$, the main contribution for $m^2$ in the MDR
\begin{equation}
    E^2-p^2+|A|m^{-n-2k} E^{n+2}(E^2-p^2)^k=m^2
\end{equation}
is $|A|m^{-n-2k} E^{n+2}(E^2-p^2)^k$, i.e.,
\begin{equation}
    |A|m^{-n-2k} E^{n+2}(E^2-p^2)^k\simeq m^2,
\end{equation}
and thus
\begin{equation}
    E^2-p^2\simeq |A|^{-\frac{1}{k}}m^{2+\frac{n+2}{k}}E^{-\frac{n+2}{k}}=m^2\left(\frac{\elvc}{E}\right)^\frac{n+2}{k}.\label{eq:sol_sub}
\end{equation}
In this case, $E^2-p^2\ll m^2$.
Combining the expansion \eq{eq:MDR_expand}, we see that as the energy gets larger, $E^2-p^2$ gets smaller and runs from $m^2$ to 0.

As the end of this section, we summarize the result of the above discussion. 
For the superluminal case ($A>0$), we have:
\begin{itemize}
\item If $|k|\ll 1$, the expansion \eq{eq:mdr_ord} works below the scale $\mathrm{min}\{\elvi, |k|^{-\frac{1}{n+2}}\elvc\}$. Especially, if $k=0$, then the MDR \eq{eq:mdrE} reduces to \eq{eq:mdr_ord} and works below $\elvi$;

\item If $-n/2\leq k<0$ and $|k|\sim O(1)$, then $\elve>\elvi>\elvc$, the MDR \eq{eq:mdrE} works below the scale $\elve$, but the expansion \eq{eq:mdr_ord} works only below $\elvc$ with $E^2-p^2=m^2+O(A)$. For $\elve\gg E\gg \elvc$, the expansion changes to \eq{eq:sol_over} with $E^2-p^2\gg m^2$. As the energy increases, $E^2-p^2$ gets larger.

\item If $0<k<1$ and $|k|\sim O(1)$, then $\elvi>\elve>\elvc$, still the MDR \eq{eq:mdrE} works below the scale $\elve$, but the expansion \eq{eq:mdr_ord} works only below $\elvc$ with $E^2-p^2=m^2+O(A)$. For $\elve\gg E\gg \elvc$, the expansion changes to \eq{eq:sol_over} and $E^2-p^2\gg m^2$. As the energy increases, $E^2-p^2$ gets larger.

\item If $k\geq 1$, then $\elvi>\elvc\geq\elve$, both the MDR \eq{eq:mdrE} and the expansion \eq{eq:mdr_ord} work below $\elve$.
\end{itemize}

For the subluminal case ($A<0$), we have:
\begin{itemize}
\item If $-n/2\leq k< 0$, though the MDR \eq{eq:mdrE} is thought to be effective below $\elve$, the energy cannot be greater than $\elvc$,
% and below this scale it can be expanded as \eq{eq:mdr_ord};
and for a given $p$ there are two solutions for $E$, one can be expressed as \eq{eq:mdr_ord} with $E^2-p^2=m^2+O(A)$, and the other can be expressed as \eq{eq:sol_unp} with $E^2-p^2\ll m^2$ and may be unphysical;
\item If $1>k>0$ and $k=O(1)$, then $\elvi>\elve>\elvc$,  the MDR \eq{eq:mdrE} works below the scale $\elve$, but the expansion \eq{eq:mdr_ord} works only below $\elvc$ with $E^2-p^2=m^2+O(A)$. For $\elve\gg E\gg \elvc$, the expansion changes to \eq{eq:sol_sub} and $E^2-p^2\ll m^2$. As the energy increases, $E^2-p^2$ gets smaller. 

\item If $k\geq 1$ or $0<k\ll 1$, the behavior is similar to that of the superluminal case.
\end{itemize}

% These MDRs derived from \eq{eq:Finsler_main} are either entirely superluminal or entirely subluminal, corresponding to $s=+1$ or $s=-1$.
% In the two cases, physics is different.

% \subsection{Case of Superluminal}

% In the superluminal case, $s=+1$.
\subsection{Physics Meaning of $\elvc$}\label{sec:4.3}

%In the previous section, we have seen that $\elvc$ appears everywhere.
%So, what is the physical meaning of $\elvc$?
%For a particle in the superluminal universe, as the energy of the particle increases, the speed of the particle increases. 
%At a certain energy level, its speed arrives at the speed of light in special relativity.
%For a particle in the subluminal universe, a case is that as the energy of the particle increases, the speed of the particle first increases, and when the energy arrives at a certain energy level, its speed decreases.
%We claim that these two energy levels are both around $\elvc$.
In the previous section, we observed that $\elvc$ is a recurring parameter. But what is the physical significance of $\elvc$?
For a particle in a superluminal universe, as the energy of the particle increases, its speed also increases. At a certain energy threshold, the particle speed reaches the speed of light as defined in special relativity.
Conversely, for a particle in a subluminal universe, the behavior is different: as the energy of the particle increases, its speed initially rises, but upon reaching a specific energy level, the speed begins to decrease.
We propose that both of these energy thresholds are approximately equal to $\elvc$.

In the superluminal case, the speed of the particle can be calculated by \eq{eq:speed_single}.
Combining the MDR \eq{eq:mdrE}, let $v=1$ and denote $S$ as $E^2-p^2$, we have
\begin{equation}
    -S + A E^{2 + n} m^{-2 k - n} (2 + k + n) S^k=0,
\end{equation}
thus
\begin{equation}
    S=(k+n+2)^{\frac{1}{1-k}}A^{\frac{1}{1-k}}  E^{\frac{n+2}{1-k}} m^{\frac{-2 k-n}{1-k}}.
\end{equation}
Combining the MDR, we can solve $E$ as
\begin{equation}
    E=C m A^{-\frac{1}{n+2}}=C \elvc,
\end{equation}
where
\begin{equation}
    C=\left((k+n+2)^{\frac{1}{1-k}}-(k+n+2)^{\frac{k}{1-k}}\right)^{\frac{k-1}{n+2}}.
\end{equation}
We can easily check that $1\leq C\leq\frac{1}{\sqrt{3}}$.
So the energy scale where the energy of the particle becomes $c$ is around $\elvc$.

In the subluminal case, we need to solve the equation $\frac{\partial v}{\partial p}=0$ together with the MDR \eq{eq:mdrE}.
Of course, there is the case that the speed of the particle increases monotonically with its energy, like $n=2$ together with $k=1$, but we focus on the case that there is a turning point.
Denote $E^2-p^2$ as $S$ and consider that $E\simeq p$,
the equation $\frac{\partial v}{\partial p}=0$ gives
\begin{eqnarray}
0=&&2 |A|^3 k E^{3 n+6} (k+n+2) (2 k+n+2) S^{3 k}\nonumber\\
&&+12 |A|^2 k E^{2 n+4} (k+n+2)  m^{2 k+n}S^{2 k+1}\nonumber\\
&&-2 |A|E^{n+2} \left(-6 k+n^2+n-2\right)  m^{2 (2 k+n)}S^{k+2}\nonumber\\
&&+4  m^{3 (2 k+n)}S^3\nonumber.
\end{eqnarray}
Let
\begin{equation}
    S=T |A|^{\frac{1}{1-k}}  m^{\frac{2 k+n}{k-1}}E^{\frac{n+2}{1-k}},
\end{equation}
we have
\begin{eqnarray}
0=&&k \left(2 k^2+3 k (n+2)+(n+2)^2\right) T^{3 k}\nonumber\\
&&+\left(6 k-n^2-n+2\right) T^{k+2}\nonumber\\
&&+6 k (k+n+2)T^{2 k+1}+2 T^3\nonumber.
\end{eqnarray}
The above equation of $T$ only contains $k$ and $n$. 
% Thus, we can assume its solution is $T_{n,k}$.
If the above equation has a positive root, we denote it as $T_{n,k}$, like $T_{2,0}=2$.
% For example, $T_{2,0}=2$.
Combining the MDR and the relation between $S$ and $T$, we have
\begin{equation}
    E=T_{n,k}^{\frac{(k-1) k}{n+2}} m A^{-\frac{1}{n+2}} =T_{n,k}^{\frac{(k-1) k}{n+2}} \elvc.
\end{equation}
So, if there is a turning point, the turning point is also around $\elvc$.

We now see the physics meaning of $\elvc$.
In the superluminal case, as the energy increases to $\elvc$, the term $Am^{-n-2k} E^{n+2}(E^2-p^2)^k$ becomes larger than $m^2$, and so do $E^2-p^2$, and the speed of the particle becomes larger than $c$.
In the subluminal case, if there is a turning point, as the energy increases to $\elvc$, the term $|A|m^{-n-2k} E^{n+2}(E^2-p^2)^k$ becomes close to $m^2$ and $E^2-p^2$ becomes close to zero, and the speed of the particle starts to decrease.

\subsection{Combining the Known Results from %Experiments
Phenomenological Analyses}\label{sec:4.4}

%Nowadays, there are many tests on LV of astroparticles based on data from spatial and ground detectors, and these tests focus on neutral particles, like photons~\cite{Shao:2009bv, Zhang:2014wpb, Xu:2016zxi, Xu:2016zsa, Zhu:2021pml} and neutrinos~\cite{Jacob:2006gn, Amelino-Camelia:2015nqa, Amelino-Camelia:2016fuh, Amelino-Camelia:2016ohi, Huang:2018ham, Huang:2019etr, Huang:2022xto, Amelino-Camelia:2022pja}.

Currently, numerous tests of Lorentz violation in astroparticles have been conducted using data from both space-based and ground-based detectors, focusing primarily on neutral particles such as photons~\cite{Shao:2009bv, Zhang:2014wpb, Xu:2016zxi, Xu:2016zsa, Zhu:2021pml,Song:2024and} and neutrinos~\cite{Jacob:2006gn, Amelino-Camelia:2015nqa, Amelino-Camelia:2016fuh, Amelino-Camelia:2016ohi, Huang:2018ham, Huang:2019etr, Huang:2022xto, Amelino-Camelia:2022pja}.

%For photons, the time-delay tests suggest that photons exhibit subluminal Lorentz violation of order one with the LV scale around $E_{\rm LV}^\gamma\simeq 3.6 \times 10^{17}\gev$ or larger.
%For neutrinos, early research based on the data from IceCube suggests both subluminal and superluminal Lorentz violation of order one with the LV scale around $E_{\rm LV}^\nu\simeq 6.4 \times 10^{17}\gev$.
%Recently, Giovanni A.-C. analyzes the new data from IceCube and suggests that the superluminal LV is %completely ruled out
%not favored~\cite{Amelino-Camelia:2022pja}.

For photons, time-delay tests indicate that they exhibit subluminal Lorentz violation of order one, with the LV scale estimated to be around $E_{\rm LV}^\gamma\simeq 3.6 \times 10^{17}\gev$ or larger. In the case of neutrinos, early research based on IceCube data suggests the presence of both subluminal and superluminal Lorentz violation, also of order one, with an LV scale around $E_{\rm LV}^\nu\simeq 6.4 \times 10^{17}\gev$. Recently, Giovanni A.-C. analyzed new IceCube data and concluded that superluminal Lorentz violation is not favored~\cite{Amelino-Camelia:2022pja}.

%Lorentz violation from the Finsler universe is either entirely subluminal or superluminal.
%Since these tests suggest that both photons and neutrinos exhibit subluminal Lorentz violation of order one, these tests are in good agreement with the physical picture of LV in the Finsler universe.
%Besides, the recent constraint for electrons of the superluminal case from LHAASO data is $E_{\mathrm{LV}e}^{(\mathrm{sup})}\gtrsim10^{26}\text{ GeV}\simeq 10^7 E_{\rm Pl}$~\cite{Li:2022ugz}, and this suggests that electrons are unlikely to be associated with superluminal Lorentz violation and is also consistent with the picture of the Finsler universe.

In the context of the Finsler universe, Lorentz violation can be categorized as either entirely subluminal or superluminal. Given that the tests indicate both photons and neutrinos exhibit subluminal Lorentz violation of order one, these findings are consistent with the physical framework of LV in the Finsler universe. Furthermore, recent constraints on electrons from LHAASO data indicate a superluminal LV scale of $E_{\mathrm{LV}e}^{(\mathrm{sup})}\gtrsim10^{26}\text{ GeV}\simeq 10^7 E_{\rm Pl}$~\cite{Li:2022ugz,He:2022jdl,He:2023ydr}, suggesting that electrons are unlikely to be associated with superluminal Lorentz violation, which is also consistent with the Finsler universe model.

%Here we use the result of neutrinos to constrain the Finsler norm \eq{eq:Finsler_main}.
%The subluminal of neutrinos suggests $A<0$, and the time delay indicates $n=1$.
%Since neutrinos have mass $m\simeq 0.1\ev$, we have
We utilize the results from neutrinos to constrain the Finsler norm given by \eq{eq:Finsler_main}. The subluminal nature of neutrinos implies ($A < 0$), and the time delay indicates ($n = 1$). Given that neutrinos have a mass of approximately $m\simeq 0.1\ev$, we can express:
\begin{equation}
    |A|=m/E_{\rm LV}^\nu\simeq 2.8\times 10^{-28},
\end{equation}
%We can see that $A$ is a very small number compared to 1, and the Finsler correction matters only when the speed of the particle is close to $c$, in other words, the energy of the particle is high enough.The critical scale for neutrinos is
This value of $A$ is significantly smaller than 1, indicating that the Finsler correction becomes relevant only when the particle speed approaches $c$, which occurs at sufficiently high energies. The critical scale for neutrinos is defined as:
\begin{equation}
    \elvc=m|A|^{-1/3}=0.15\gev,
\end{equation}
%and indeed $\elvc\ll\elvi=E_{\rm LV}^\nu$.
and indeed, we find that $\elvc\ll\elvi=E_{\rm LV}^\nu$.

%We now constrain the $k$ in the MDR of neutrinos.
%If $k<0$, the energy of neutrinos cannot reach $\elvc=0.15\gev$, and it is impossible.
%If $k\geq 0$, the MDR works below $\elve=m |A|^{-\frac{1}{1+2k}}$, which means there might be new physics around $\elve$.
%Since we have observed neutrinos with energy reaching PeV and have not found new physics in the PeV region, we may have $\elve \gg 1\pev$, or $0\leq k<0.36$.
%If $k$ is not close to zero, then in the $\elvc\ll E\ll \elve$ region, the MDR of neutrinos will be expanded as \eq{eq:sol_sub}.
%If we want that the expansion \eq{eq:MDR_expand} holds for PeV neutrinos, then $k^{-\frac{1}{3}}\elvc\gg 1\pev$ and $k<3.6\times 10^{-21}$.
%In this scenario, it is likely that $k$ equals zero.

Next, we constrain the parameter $k$ in the modified dispersion relation (MDR) for neutrinos. If $k<0$, then neutrinos cannot reach the energy scale $\elvc=0.15\gev$, which is impossible. If $k\geq 0$, the MDR is valid below $\elve=m |A|^{-\frac{1}{1+2k}}$, indicating that new physics may emerge around this scale. Given that we have observed neutrinos with energies reaching the PeV range without encountering new physics in that region, we infer that $\elve \gg 1\pev$, or equivalently, $0\leq k<0.36$.

If $k$ is not close to zero, then in the region where $\elvc\ll E\ll \elve$, the MDR for neutrinos can be expanded as given in \eq{eq:sol_sub}. To ensure that the expansion \eq{eq:MDR_expand} holds for PeV neutrinos, we require $k^{-\frac{1}{3}}\elvc\gg 1\pev$ and $k<3.6\times 10^{-21}$. Under these conditions, it is likely that $k$ is approximately zero.

%In the picture of the Finsler universe, we can predict the LV scale of other massive particles.
%For electrons, its LV scale $\elvi^{\rm e}=m_{\rm{e}}/A=1.8\times 10^{24}\gev\simeq 10^5 E_{\rm Pl}$.
% So it is difficult to test the LV effects of electrons, 
%This prediction is consistent with the results of Ref.~\cite{Maccione:2007yc}, in which the constraint for electrons from the Crab Nebula in the subluminal case is $E_{\mathrm{LV}e}^{(\mathrm{sub})}\gtrsim10^{24}\text{ GeV}$  at 95\% confidence.

In the framework of the Finsler universe, we can also predict the LV scale for other massive particles. For electrons, the predicted LV scale is given by $E_{\rm LV,I}^{\rm e}=m_{\rm{e}}/A=1.8\times 10^{24}\gev\simeq 10^5 E_{\rm Pl}$. This prediction aligns with the results from Ref.~\cite{Maccione:2007yc}, where the constraint for electrons from the Crab Nebula in the subluminal case is found to be $E_{\mathrm{LV}e}^{(\mathrm{sub})}\gtrsim10^{24}\text{ GeV}$ at a 95\% confidence level.

\section{Summary and Discussion}\label{sec:5}

%Finsler geometry provides a good tool to explore the spacetime of Lorentz violation.
%In previous research, the Finsler structures of different particles are considered to be distinct.
%In this work, based on several considerations in Sec.~\ref{sec:3}, we propose the hypothesis that the Finsler structure is the property of the universe.
%In this hypothesis, the universe is either subluminal ($A<0$) or superluminal ($A>0$). 
%That means that there cannot exist both subluminal and superluminal Lorentz violations in the universe, which matches well with the test of photons and neutrinos with subluminal LV discussed in Sec.~\ref{sec:4.4}.

Finsler geometry serves as a powerful framework for exploring the spacetime associated with Lorentz violation. In previous research, the Finsler structures attributed to different particles were considered distinct. In this work, based on several considerations discussed in Sec.~\ref{sec:3}, we propose the hypothesis that the Finsler structure is an intrinsic property of the universe. According to this hypothesis, the universe can be categorized as either subluminal ($A<0$) or superluminal ($A>0$). This implies that both subluminal and superluminal Lorentz violations cannot coexist within the same universe, which aligns well with the phenomenological results concerning photons and neutrinos exhibiting subluminal LV, as discussed in Sec.~\ref{sec:4.4}.

%In the picture of the Finsler universe, the time-delay formula depends on the $n$ and $A$ of the Finsler norm \eq{eq:Finsler_main}, so it is the same for all particles. 
%However, from the Finsler norm, we find a series of solutions of MDRs. 
%To study the behavior of different MDRs of massive particles, we introduce three different LV scales as $\elvi=m|A|^{-\frac{1}{n}}$, $\elve=m|A|^{-\frac{1}{n+2k}}$ and $\elvc=m|A|^{-\frac{1}{n+2}}$.
%We can see that these three LV scales are all proportional to the mass of the particle. $\elvi$ plays a role in the experiments of time delay, $\elve$ determines the energy scale at which the MDR is valid, and $\elvc$ determines the critical point at which the expansion \eq{eq:mdr_ord} becomes invalid.
%Detailed discussions of the behavior of different MDRs in the subluminal or superluminal universe can be found in Sec.~\ref{sec:4.2}.
%In Sec.~\ref{sec:4.4}, we show that $\elvc$ is related to the behavior of the speed of the particle. In the superluminal universe, $\elvc$ is the scale where the speed arrives $c$, and in the subluminal universe, $\elvc$ is around the turning point of the speed, if the turning point exists.

Within the framework of the Finsler universe, the time-delay formula is dependent on the parameters $n$ and $A$ of the Finsler norm \eq{eq:Finsler_main}, suggesting that it is uniform across all particles. However, the Finsler norm leads us to a series of solutions for modified dispersion relations (MDRs). To analyze the behavior of these MDRs for massive particles, we introduce three distinct LV scales: 
$\elvi=m|A|^{-\frac{1}{n}}$, $\elve=m|A|^{-\frac{1}{n+2k}}$ and $\elvc=m|A|^{-\frac{1}{n+2}}$. Notably, all three LV scales are proportional to the mass of the particle. The scale $\elvi$ plays a significant role in time-delay experiments, while $\elve$ defines the energy scale at which the MDR is applicable. The scale $\elvc$ indicates the critical point beyond which the expansion \eq{eq:mdr_ord} becomes invalid. A detailed discussion of the behavior of different MDRs in subluminal and superluminal universes can be found in Sec.~\ref{sec:4.2}. In Sec.~\ref{sec:4.4}, we demonstrate that $\elvc$ is related to the particle's speed behavior. In a superluminal universe, $\elvc$ marks the scale at which the speed approaches $c$, while in a subluminal universe, $\elvc$ corresponds to the turning point of the speed, if such a turning point exists.

%Combining the tests of LV from IceCube neutrinos, we find that the parameter of the Finsler universe is $A\simeq-10^{-28}$.
%For the MDR of neutrinos, if $k\neq 0$, then above the scale $\elvc=0.15\gev$, the MDR can be expanded as \eq{eq:sol_sub}, quite different from the ordinary one.
%The picture of the Finsler universe makes predictions for other particles, since in this picture the LV scales are proportional to the masses of particles.
%For example, the LV scale for electrons is predicted as $\elvi^{\rm e}\simeq 1.8\times 10^{24}\gev$, about $10^5$ times the Plank scale.
%Such a large LV scale makes it difficult to test LV effects for electrons and is consistent with the test of LV of electrons~\cite{Maccione:2007yc}.
%For heavier particles, we will never detect their LV effects easily.
%Though the Finsler geometry cannot describe messless particles with LV directly, we can see the trend that the LV scales increase as the masses of different particles increase. 
%This is true for current results of photons ($\elvi^\gamma\simeq 3.6 \times 10^{17}\gev$) and neutrinos ($\elvi^\nu\simeq 6.4 \times 10^{17}\gev$).

By combining the tests of LV from IceCube neutrinos, we estimate the parameter of the Finsler universe to be $A\simeq-10^{-28}$. For the MDR of neutrinos, if 
$k\neq 0$, then above the scale 
$\elvc=0.15\gev$, the MDR can be expressed as \eq{eq:sol_sub}, which differs significantly from the conventional form. The Finsler universe framework also makes predictions for other particles, as the LV scales are proportional to their masses. For instance, the LV scale for electrons is predicted to be 
$E_{\rm LV,I}^{\rm e}\simeq 1.8\times 10^{24}\gev$, approximately $10^5$ times the Planck scale. Such a large LV scale presents challenges for testing LV effects in electrons and is consistent with existing tests of LV in electrons~\cite{Maccione:2007yc}. 
%For heavier particles, detecting LV effects remains exceedingly difficult. 
Although Finsler geometry cannot directly describe massless particles with LV, it reveals a trend where LV scales increase with the mass of different particles. This trend is supported by current constraints %results 
for photons ($E_{\rm LV,I}^\gamma\simeq 3.6 \times 10^{17}\gev$) and neutrinos ($E_{\rm LV,I}^\nu\simeq 6.4 \times 10^{17}\gev$).

%Assuming it to be an effective theory to describe the spacetime of LV, the picture of the Finsler universe exhibits rich content.
%In this picture, the most important result is that the LV scales of different massive particles are proportional to their masses.
%Also, a series of MDRs arise in the Finsler universe, and their properties differ significantly from those typically studied.
%Although it lacks the ability to describe massless particles with Lorentz violation, it still provides us with a glimpse into the realm of the spacetime of quantum gravity.

As an effective theory for describing LV spacetime, the Finsler universe presents a rich tapestry of phenomena. The most significant outcome of this framework is the observation that the LV scales of various massive particles are proportional to their masses. Furthermore, a series of MDRs emerge within the Finsler universe, with properties that diverge notably from those typically studied. Although this framework lacks the capacity to directly address massless particles exhibiting Lorentz violation, it offers valuable insights into the nature of spacetime in the context of quantum gravity.

% \section*{Acknowledgements}

% This work is supported by
\section*{Acknowledgements}
This work was supported in part by the National Natural Science Foundation of China under Grant No.~12547101. HL was also supported by the start-up fund of Chongqing University under No.~0233005203009, and JZ was supported by the start-up fund of Chongqing University under No.~0233005203006.

\bibliography{apssamp}% Produces the bibliography via BibTeX.

\end{document}